# Doppler peaks and all that:
# CMB anisotropies and what they can tell us

Max Tegmark †

Max-Planck-Institut für Physik, Föhringer Ring 6,D-80805 München
Max-Planck-Institut für Astrophysik, Karl-Schwarzschild-Str. 1, D-85740 Garching
max@mppmu.mpg.de

**Abstract.** The power spectrum of fluctuations in the cosmic microwave background (CMB) depends on most of the key cosmological parameters. Accurate future measurements of this power spectrum might therefore allow us to determine $h$, $\Omega$, $\Omega_b$, $\Lambda$, $n$, $T/S$, *etc.*, with hitherto unprecedented accuracy. In these lecture notes, we review the various physical processes that generate CMB fluctuations, focusing on how changes in the parameters alters the shape of the power spectrum. We also discuss foregrounds and real-world data analysis issues and how these affect the accuracy with which the parameters can be measured.

## 1. Introduction

These are exciting times for CMB researchers. The first detection of CMB fluctuations (other than the kinematic dipole) were announced only three years ago [1], shortly before the 1992 Varenna cosmology summer school. Since then, the field has virtually exploded with activity, on both the experimental and theoretical fronts. On the experimental side, ground- and balloon-based experiments have now produced more than a dozen independent detections of fluctuations, over a range of angular scales, and many more experiments are currently under way or being planned for the near future — see [2–4] for recent reviews. On the theoretical side, great progress has been made both in understanding and quantifying the various physical effects that generate anisotropies (*e.g.* [5–10]) and in developing data-analysis techniques that enable us to extract cosmological information from real-world CMB data (*e.g.* [11–19]). It now seems plausible that that the next generation of CMB experiments, together with these theoretical advances, will allow accurate determination of many of the key cosmological parameters.

---

† To appear in Proc. Enrico Fermi, Course CXXXII, Varenna, 1995.
Available from *http://www.mpa-garching.mpg.de/~max/varenna.html* (faster from Europe)
and from from *http://astro.berkeley.edu/~max/varenna.html* (faster from the US).

Section 2 of the these lecture notes is an overview of the main physical processes that are produce anisotropies in the microwave sky, including foregrounds. Section 3 gives a more detailed discussion of the so-called primary anisotropies, focusing on how changes in cosmological parameters alter the power spectrum in Figure 1. In section 4 we discuss how accurately these parameters can be measured given real-world problems such as pixel-noise, beam dilution, foreground contamination and incomplete sky coverage. Finally, Appendix A contains a brief primer on power spectra, spherical harmonics, *etc.*

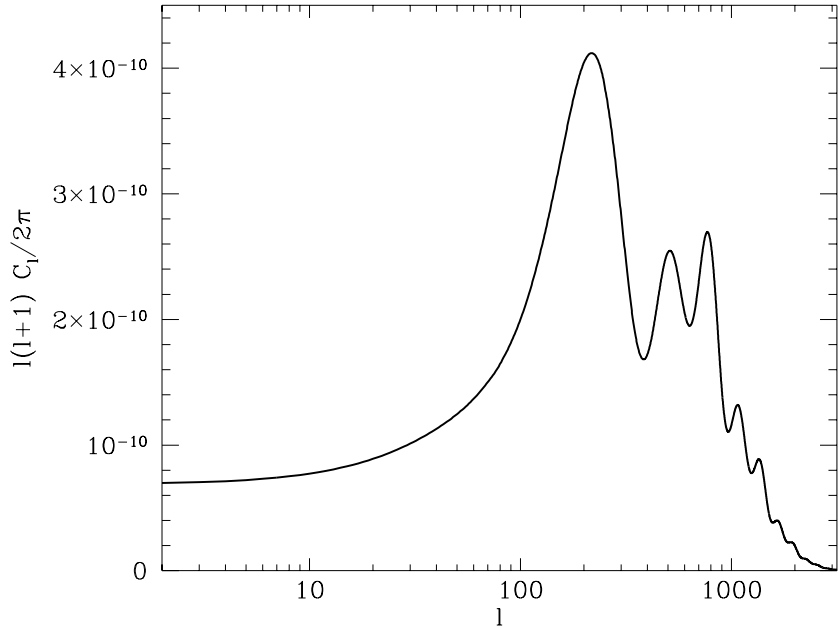

**Figure 1.** The angular power spectrum of the standard CDM model [20]($\Omega = 1$, $\Omega_b = 0.06$, $h = 0.5$, $n = 1$).

## 2. Sources of CMB fluctuations: an overview

We will first give a brief summary of all the main effects, then return to a more detailed discussion of the primary anisotropies and their dependence on cosmological parameters in the following section.

There are many different ways in which one may chose to categorize the various physical processes that affect the temperature distribution we measure in the microwave sky. One common approach (*e.g.* [21]) is to order them chronologically, according to how far back in time they occurred, and this is roughly the approach taken in Table 1.

**Table 1.** Sources of temperature fluctuations.

| | | |
|---|---|---|
| **PRIMARY** | Gravity | |
| | Doppler | |
| | Density fluctuations | |
| | Damping | |
| | Defects | Strings |
| | | Textures |
| **SECONDARY** | Gravity | Early ISW |
| | | Late ISW |
| | | Rees-Sciama |
| | | Lensing |
| | Local reionization | Thermal SZ |
| | | Kinematic SZ |
| | Global reionization | Suppression |
| | | New Doppler |
| | | Vishniac |
| **"TERTIARY"** | Extragalactic | Radio point sources |
| | | IR point sources |
| (foregrounds | Galactic | Dust |
| & | | Free-free |
| headaches) | | Synchrotron |
| | Local | Solar system |
| | | Atmosphere |
| | | Noise, *etc.* |

## 2.1. Primary fluctuations

As the CMB photons decouple from the baryons around a redshift $z \sim 10^3$, about $6000h^{-1}$Mpc away from us, they take with them three different imprints of the region from which they last scatter, corresponding to the peculiar gravitational potential $\phi$, the radial peculiar velocity $v_r$ and density fluctuation $\delta$:

(i) Photons last scattered in a potential well ($\phi < 0$) will experience a gravitational redshift as they climb out of it.

(ii) Photons last scattered by matter whose peculiar velocity is away from us ($v_r > 0$) will receive a Doppler redshift.

(iii) Photons emanating from an overdense region ($\delta > 0$) will have a higher temperature, simply because denser regions are intrinsically hotter.

These three effects correspond to the first three entries in Table 1, and are summarized by the equation

$$\frac{\Delta T}{T}(\hat{\mathbf{r}}) = \phi(\mathbf{r}) - \hat{\mathbf{r}} \cdot \mathbf{v}(\mathbf{r}) + \frac{1}{3}\delta(\mathbf{r}), \tag{1}$$

where the length of the vector $\mathbf{r}$ is the comoving distance to the surface of last scattering, *i.e.*, $r \approx 6000h^{-1}$Mpc, and the fields $\phi$, $\mathbf{v}$ and $\delta$ are to be evaluated at the time of recombination, at $z \sim 10^3$. Here and throughout, we use units where the speed of light $c = 1$. As we will discuss further on, adiabatic initial conditions lead to a situation where the locations of the overdensities coincide with those of the potential wells, so that the third term partially cancels the first. It turns out that for fluctuations on very large scales ($\gg$ a few degrees), $\delta \approx -2\phi$, so that these two terms combine to simply $\phi/3$, the so-called Sachs-Wolfe effect [22], which is responsible for the flat part on the left side in Figure 1. On smaller scales, the fluctuations in $\phi$, $\mathbf{v}$ and $\delta$ have time to undergo a type of acoustic oscillations before recombination occurs, and as we will see further on, this is responsible for the so-called Doppler peaks further to the right in Figure 1.

Since the last-scattering surface (LSS) is not infinitesimally thin, the CMB temperature that we measure in a given direction on the sky will in fact be a weighted average, corresponding to a mixture of photons coming from the near and far parts of the LSS. This effect, listed as the fourth in Table 1 as a "negative source", effectively washes out fluctuations on scales below that corresponding to the LSS thickness, which corresponds to about $\ell \sim 10^3$ ($\theta \sim 0.1°$) – see Figure 1. This is quite analogous to the way in which smoothing with a wide experimental beam washes out small-scale fluctuations, the only difference being that the latter smoothing is in the transverse rather than in the radial direction.

We will return to these primary sources of fluctuations in section 2. Alternative scenarios based on topological defects are not discussed in these notes (the interested reader is referred to [23–31] and references therein), nor will tensor mode fluctuations be covered — for $C_\ell$-formulas for this case, see *e.g.* [32,33] and references therein.

### 2.2. Secondary fluctuations

The effects listed under this heading in Table 1 roughly speaking refer to processes that affect the CMB photons on their way from the last scattering surface to us, *i.e.*, in the redshift range $0 < z < 10^3$.

*2.2.1. Gravitational effects*

After recombination, the baryons have essentially lost their ability to interact with the photons through Thomson scattering, but will of course continue to affect the photons gravitationally just as they did before. These gravitational effects are conveniently divided into four categories, as in Table 1. The first three (the Early ISW, Late IWS and Rees-Sciama effects) are all manifestations of the so called Integrated Sachs-Wolfe (ISW) effect [22], whereby time-variation in the gravitational potential imprints temperature fluctuations on the CMB. The ISW effect on a single photon is conveniently written as an integral along its flight path:

$$\frac{\Delta T}{T} = \int \dot{\phi}[\mathbf{r}(t), t] dt, \qquad (2)$$

where $\dot{\phi}$ is the conformal time derivative of $\phi$ at a fixed position. This has a simple physical interpretation. If a photon flies through a potential well, the blueshift it acquires when falling in will be exactly canceled by its redshift from climbing out. If the potential well becomes more shallow while the photon is in it ($\dot{\phi} > 0$), the cancellation will no longer be perfect, and a net blueshift will result.

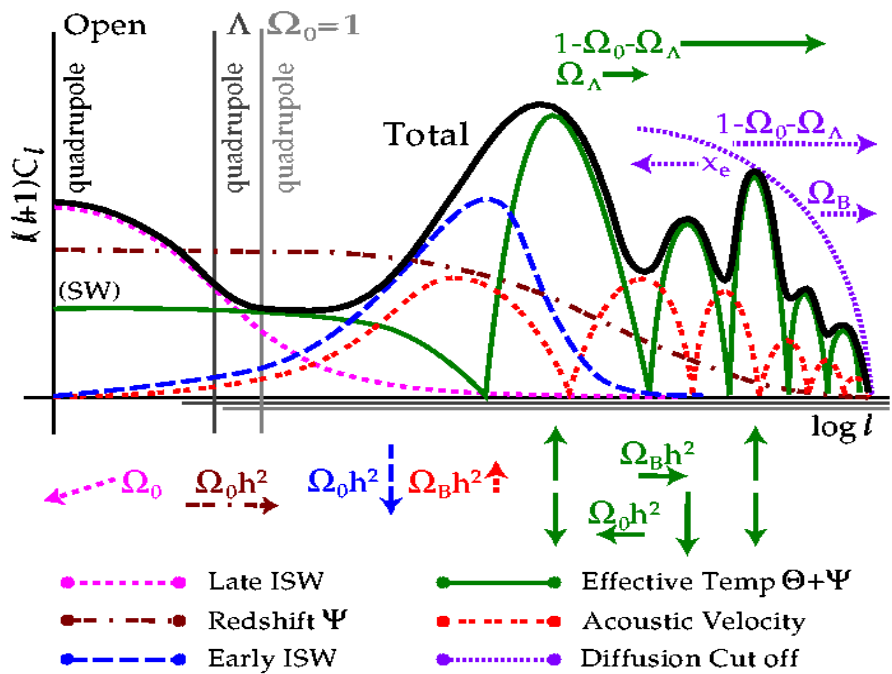

**Figure 2.** Primary and ISW contributions to the CDM power spectrum. Reprinted from [9].

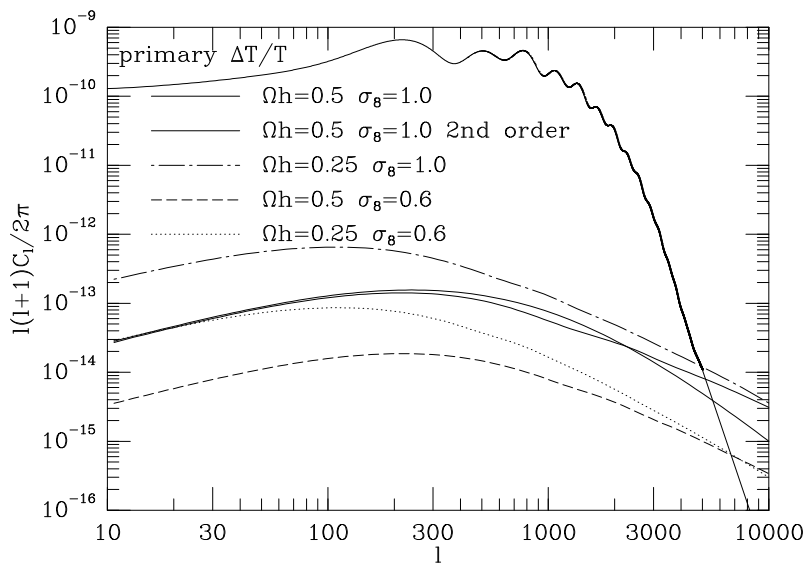

**Figure 3.** The Rees-Sciama effect. Reprinted from [36].

In the linear regime after matter-domination, it is well-known that the peculiar gravitational potential $\phi$ remains constant, in which case there is no ISW effect. There are three different cases when $\dot{\phi} \neq 0$, and the corresponding contributions to the ISW effect are usually called the early ISW, late ISW and Rees-Sciama effects, respectively.

(i) Shortly after recombination, the photon contribution to the density of the universe is still not altogether negligible. As a result, $\phi$ decays somewhat, causing the **early ISW effect**. Since the photon density is fixed by the CMB temperature today, this effect becomes more important when $\Omega$, the total density, is lowered.

(ii) If $\Lambda > 0$, the universe will eventually become vacuum dominated. If $\Omega + \Lambda \neq 1$, the universe may become curvature dominated. In both these cases, $\dot{\phi} \neq 0$. Since vacuum energy and curvature become important only at low redshifts, this is known as the **late ISW effect**.

(iii) Once non-linear structures such as galaxy clusters form, linear perturbation theory of course breaks down, and the perturbation theory result $\dot{\phi} = 0$ is no longer correct. This effect is known as the **Rees-Sciama effect** [34].

The early ISW effect is schematically illustrated in Figure 2, and typically peaks slightly to the left of the first Doppler peak. The late ISW, on the other hand, tends to peak on the very largest scales, since it becomes effective only at late times [35]. In both cases, the effects drops off on scales much smaller than the horizon scale at their formation epoch, because of $\sqrt{N}$-type cancellation effects. The contribution to $C_\ell$ from the Rees-Sciama effect is shown in Figure 3, reprinted from [36]. Although the magnitude of the Rees-Sciama effect has been subject to a long debate (*e.g.* [37–39]), it now seems as though for standard CDM models, its impact on the power spectrum is for all practical purposes negligible, being several orders of magnitude weaker than the primary anisotropies [36].

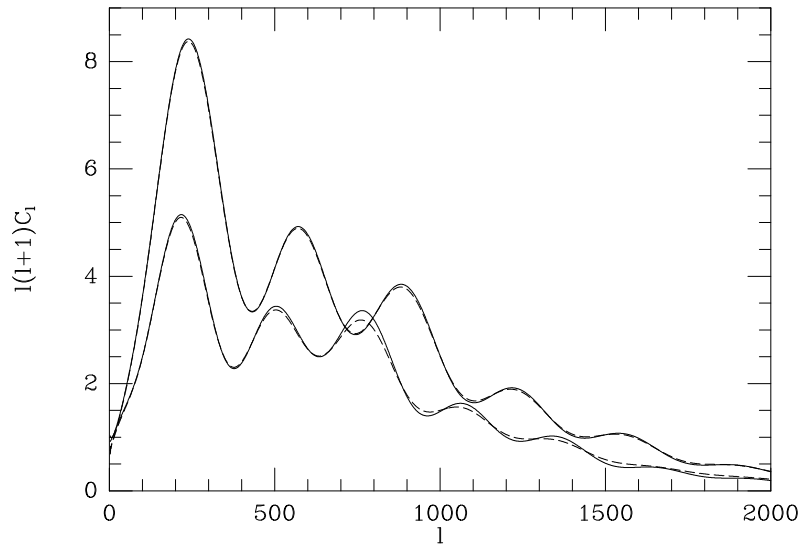

**Figure 4.** Angular power spectra including (dotted lines) and excluding (solid lines) the effect of gravitational lensing. Reprinted from [40].

The ISW effect can be thought of as the photons receiving momentum kicks from the gravitational field gradient $\nabla\phi$ parallel to their flight path. These kicks thus change the energy of the photons, but not their direction. There is also a twin effect, caused by momentum kicks from $\nabla\phi$ perpendicular to the flight path. To first order, this effect will leave the energy of the photon unchanged, but deflect its trajectory [41]. A recent evaluation of this intensely debated effect is shown in Figure 4, which is reprinted from [40]. If a pair of photons would have arrived here separated by an angle $\theta$ in the absence of fluctuations in $\phi$, then they will in reality arrive separated by some angle $\theta+\Delta\theta$. It has been shown (see [40] and references therein) that for typical models, $\Delta\theta/\theta \ll 1$, typically between 0.1 and 0.2. This means that this is essentially an issue of weak **gravitational lensing**. If we imagine the CMB fluctuations for $\phi = 0$ painted on the inside of a rubber sphere, this lensing effect will correspond to stretching and compressing the rubber in a random fashion, much like the way our mirror images get distorted by non-flat mirrors in amusement parks. Since $\Delta\theta/\theta \ll 1$, this distortion will always constitute a one-to-one mapping of the image, *i.e.*, there will be no caustics where the image "folds over" on itself. There is thus no smearing involved, so the total power in the fluctuations is conserved. Rather, as Figure 4 illustrates for two CDM models, the effect of this angular jumbling is to smear out the power spectrum somewhat, redistributing power from the peaks to the troughs. Although this effect is small (typically a few percent), it may well be detectable in future CMB experiments.

*2.2.2. Effects of local reionization*

Above, we discussed the effects of $\phi$ on the CMB photons as they free-stream to us after recombination. The other two fields, **v** and $\delta$, can clearly only influence the CMB photons if the baryons become reionized. This may happen locally, confined to for instance hot clusters of galaxies, or globally, throughout all of space.

Local reionization manifests itself in two ways, corresponding to the impact of **v** and $\delta$, respectively. Both are known as the Sunyaev-Zeldovich (SZ) effect [42–44]:

(i)  If a cluster of galaxies is moving towards us, Thomson scattering of CMB photons off of free electrons in the hot intra-cluster gas will cause a Doppler blueshift in the direction of the cluster, known as the **kinematic SZ-effect**.

(ii) Independent of the cluster velocity, the high temperature of the free electrons will distort the Planck spectrum by depleting the Rayleigh-Jeans (low $\nu$) tail and overpopulating the Wien (high $\nu$) tail. This is known as the **thermal SZ-effect**, and appears as a redshift below 218 GHz and as a blueshift above 218 GHz.

Both of these effects are likely to have a negligible impact on the overall CMB power spectrum [45]. However, they can be quite large in the directions of cluster cores, and thus be used to used to learn more about both internal properties of clusters (see *e.g.* [46–48]) and about cluster peculiar velocities (*e.g.* [49,50]).

*2.2.3. Effects of global reionization*

If reionization is global, throughout space, the effects on the CMB power spectrum can be quite radical, as seen in Figure 5, suppressing fluctuations on small scales. This is a simple geometrical effect, illustrated in Figure 6. In the coordinates of this figure, light rays propagate along 45° lines just as in Euclidean space. The horizontal circles are labeled with their corresponding redshifts, the dotted circle corresponding to the big bang. If we (at the apex of the cone) detect a CMB photon arriving from the right in the figure, then it would normally have been in the lower right corner at recombination at $z = 10^3$. However, if reionization caused the photon to Thomson scatter off of free electrons, with the last scattering occurring at $z = 10$, then the photon could have come from anywhere in the double-shaded region. In other words, the temperature we see in a given direction of the sky is really the weighted average of the temperature of part of the $z = 1000$ last scattering surface. This smearing will be on an angular scale

$$\theta \approx \sqrt{\frac{\Omega_0}{z}}, \tag{3}$$

corresponding to the angle subtended by the horizon ($\sim$ the backward light cone in Figure 6) at the redshift $z$ of last scattering.

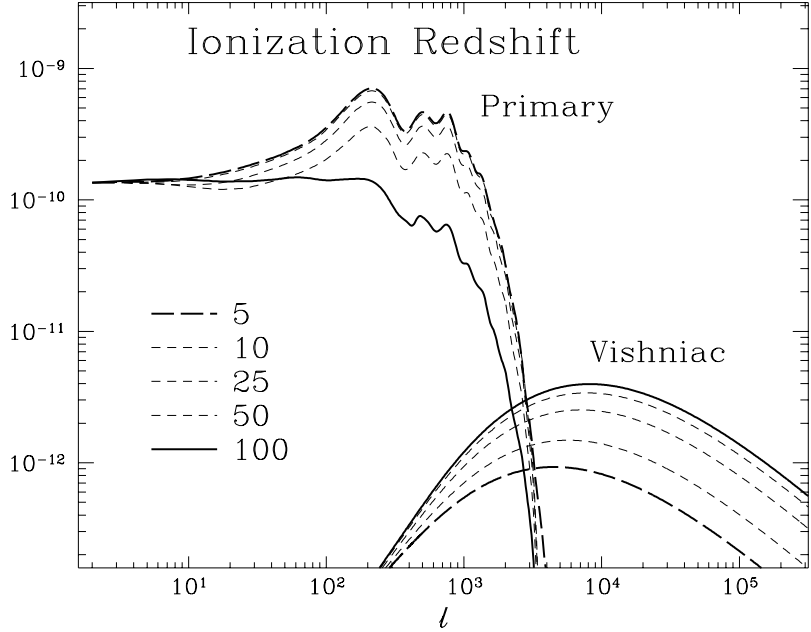

**Figure 5.** The effect of reionization in standard CDM. The curves correspond to models where the universe becomes suddenly ionized at the redshifts indicated and remains ionizaed until today. Reprinted from [9].

If the universe became reionized at a redshift $z$ and remained ionized until today, the probability that a CMB photon would never get scattered is $e^{-\tau}$, where the optical depth is (*e.g.* [51])

$$\tau \approx \Omega_0^{-1/2} \left(\frac{h\Omega_b}{0.06}\right) \left(\frac{z}{92}\right)^{3/2}. \tag{4}$$

Thus a fraction $e^{-\tau}$ of the photons remain unaffected by the reionization. Since the power is the square of the fluctuation amplitude, it will get suppressed by a factor $e^{-2\tau}$.

For standard CDM parameters, equation (4) means that even if $\tau \gg 1$, so that almost all photons got scattered at some point, the last scattering event for most photons will be around $z \sim 50$, corresponding to a maximum smearing scale $\theta \sim 8°$.

In summary, global reionization will affect the power spectrum as follows:

(i) The power on small scales ($\ell \gg 10$) will be suppressed by a factor $e^{-2\tau}$

(ii) The power on larger scales will remain unaffected.

This is clearly illustrated in Figure 5. The solid curve in the figure reveals an additional effect: around the critical scale $\theta$, extreme reionization causes a slight increase in power. This corresponds to new "primary" anisotropies from the new surface of last scattering, and will exceed the old primary fluctuations because the fluctuations **v** and $\delta$ have grown since $z = 10^3$. Finally, the so-called Vishniac effect [9,52–56] is seen to generate new fluctuations on very small scales. This is a nonlinear (second-order) effect, related to coupling between large-scale fluctuations in $v_r$ and small-scale fluctuations in $\delta$.

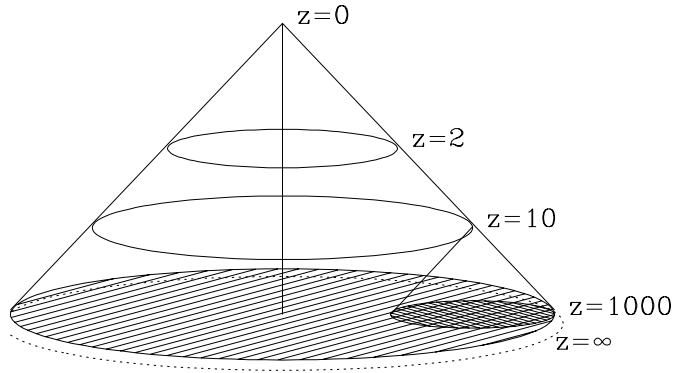

**Figure 6.** Our backward light cone in a flat $\Omega = 1$ universe is shown in comoving coordinates, with conformal time $\eta$ on the vertical axis. One of the three spatial dimensions has been suppressed.

### 2.3. "Tertiary" fluctuations (foregrounds and headaches)

Although the fluctuation sources listed as "tertiary" in Table 1 are of course not CMB fluctuations in the conventional sense, reliable parameter estimation from CMB data requires accurate knowledge of their properties.

The frequency-dependence of the various foregrounds has been extensively studied, in both "clean" and "dirty" regions of the sky (see *e.g.* [57,58] for recent reviews). However, these properties alone provide a description of the foregrounds that is somewhat too crude to assess the extent to which they can be separated from the underlying CMB signal, since the foreground fluctuations depend on the multipole moment $\ell$ as well (see [48] for simulations). Most published plots comparing different CMB experiments tend to show $\ell$ on the horizontal axis and an amplitude ($C_\ell$ or an r.m.s. $\Delta T/T$) on the vertical axis, whereas most plots comparing different foregrounds show amplitude plotted against frequency $\nu$. Since the fluctuations in the latter tend to depend strongly on both $\ell$ and $\nu$, *i.e.*, on both spatial and temporal frequency, one obtains a more accurate picture by combining both of these pieces of information and working in a two-dimensional plane as in Figure 7. Indeed, this $\ell - \nu$ plane arises naturally in a minimum-variance subtraction scheme employing multiple frequencies [19]. This figure also shows roughly the regions in this plane probed by various CMB experiments. Each rectangle corresponds to one experiment. Its extent in the $\ell$-direction shows the customary FWHM width of the experimental window function (see e.g. [59]), whereas the vertical extent shows the frequency range that is covered. For single-channel experiments, we plot the quoted bandwidth, whereas for multi-channel experiments, the box has simply been plotted in the range between the lowest and highest frequency channel. For a more detailed description of these experiments, see [3] and references therein. The various shaded areas correspond to the regions where fluctuations from various foreground sources become comparable to those of the CMB, and will be discussed in greated detail below. Comparing these with the locations of the various boxes, it is easy to understand which experiments are the most affected by the various foregrounds.

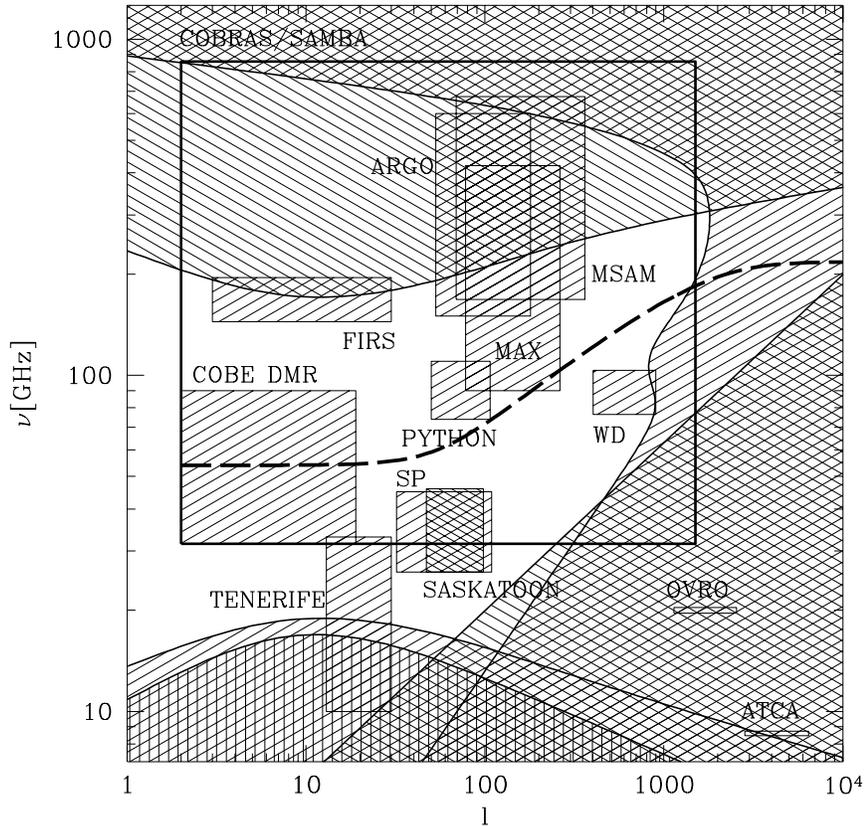

**Figure 7.** Where the various foregrounds dominate. The shaded regions indicate where the various foregrounds cause fluctuations exceeding those of COBE-normalized scale-invariant fluctuations, thus posing a substantial challenge to estimation of genuine CMB fluctuations. They correspond to dust (top), free-free emission (lower left), synchrotron radiation (lower left, vertically shaded), radio point sources (lower right) and COBRAS/SAMBA instrumental noise and beam dilution (right), all for the cleanest 20% of the sky. The heavy dashed line shows the frequency where the total foreground contribution to each multipole is minimal. The boxes indicate roughly the range of multipoles $\ell$ and frequencies $\nu$ probed by various CMB experiments. Reprinted from [19].

The foreground estimates given below are all based on [19]. They are *not* intended to be very accurate, especially when it comes to normalization. Rather, the emphasis is on their *qualitative* features, especially those that differentiate them from one another. Despite the fact that we currently lack accurate high-resolution data in many important frequency bands, we will see that quite robust qualitative conclusions can be drawn about which regions of the $\ell - \nu$-plane will be most suitable for estimating various parts of the CMB power spectrum.

### 2.4. Extragalactic point sources

In this section, we will make estimates of the angular power spectrum $C_\ell(\nu)$ for point sources. Here the $\ell$-dependence is well known, but the $\nu$-depencence quite uncertain. However, despite these uncertainties, we will see that radio point sources will contribute mainly to the lower right corner of Figure 7, whereas infrared point sources will contribute mainly to the upper right.

It is easy to show [19] that, apart from the monopole, the power spectrum of unresolved point sources is

$$C_\ell = \int_0^{\phi_c} \frac{\partial \bar{n}}{\partial \phi} \phi^2 d\phi, \qquad (5)$$

where $\frac{\partial \bar{n}}{\partial \phi}$ is the *differential source count*. In other words, we have defined $\bar{n}(\phi)$ as the number density per steradian of sources with flux less than $\phi$. Note that although $C_\ell$ will depend on the frequency $\nu$ of observation (since $\phi$ does), it is independent of $\ell$. This corresponds to a so-called white noise power spectrum, and will always dominate over a "flat" $\ell^{-2}$ CMB power spectrum at large enough $\ell$.

In real life, we are of course far from powerless against point sources, and can either attempt to subtract them by using spectral information from point source catalogues, or simply choose to throw away all pixels containing a bright point source. In either case, the end result would be that we eliminate all sources with a flux exceeding some flux cut $\phi_c$, which then becomes the upper limit of integration in equation (5). Obviously, one can never eliminate *all* point sources, as there are for all practical purposes infinitely many of them.

*2.4.1. Radio sources*

The frequency dependence of the point source counts is unfortunately very poorly known in the microwave region of the spectrum. Based on data from the 1.5 GHz

VLA FIRST point source catalog, and extrapolating to higher frequencies with a power law $B(\nu) \propto \nu^{-\alpha}$, the power spectrum of unresolved radio sources is estimated to be [19]

$$\left[\frac{2\ell+1}{4\pi}\ell C_\ell(\nu)\right]^{1/2} \approx 0.30\text{K} \left(\frac{\sinh^2(x/2)}{[\nu/1.5\text{ GHz}]^{4+\alpha}}\right)\ell, \tag{6}$$

where $x = h\nu/kT_0$. This corresponds to a 100 mJy flux cut at 1.5 GHz (for an all-sky survey, this would mean removing about 70000 sources). In Figure 7, the rather cautious assumption of an effective spectral index $\alpha = 0.0$ for the population as a whole has been used. Flat-spectrum sources with spectral index $\alpha \approx 0.3$ are likely to dominate at higher frequencies [61], but this is of course only to be used as a crude first approximation, as the emission at higher frequencies is likely to be dominated by sources whose spectra rise and peak near those frequencies, and very little is currently known about the abundances of such objects — accurate data at 50 GHz and above is badly needed.

*2.4.2. Infra-red sources*

Our observational knowledge is no better at the high frequency end, where infrared emission from high redshift galaxies could play an important role. For instance, if this emission is dominated by dust in these galaxies with emissivity $\beta = 2$ (see the following section), we would expect $\alpha = -4$ to be a better description at the higher microwave frequencies. Unfortunately, the differential source counts of such infrared point sources around 100 GHz is still completely unknown. For recent reviews of these issues, see [62,63].

### 2.5. Diffuse Galactic sources

In this section, we discuss the qualitative features we expect for the angular power spectra of the diffuse Galactic contaminants, namely dust, free-free emission and synchrotron radiation.

*2.5.1. Power spectrum*

The power spectrum of Galactic dust has been estimated by a number of authors [64–66,19], using the IRAS all-sky survey [67] and the COBE/FIRAS data. The 100 micron IRAS map has an angular resolution of two arcminutes, *i.e.*, better than most proposed CMB satellites would attain. Although the amplitude varies greatly with Galactic latitude, the overall shape is strikingly independent of latitude, and typically declines as $C_\ell \propto 1/\ell^3$ for $\ell$ between 100 and a few thousand, steepening slightly on the smallest scales.

Direct estimates of the power spectrum of synchrotron radiation are based on the Haslam 408 GHz map [68]. Although the angular resolution of this map is only of order 0.85°, *i.e.*, far too low to provide information for $\ell \gg 100$, the logarithmic slope has been found to be consistent with that for dust in the overlapping multipole range; around $-3$ [66,19]. These results are hardly surprising, since even without analyzing observational data, one may be able to guess the qualitative features of the power spectra of the three diffuse components. Since they are all caused by emission from diffuse blobs, one might expect their power spectra to exhibit the following characteristic features:

(i)  $C_\ell$ independent of $\ell$ for small $\ell$, corresponding to scales much greater than the coherence length of the blobs (this is the standard Poisson behavior, and follows if one assumes that well separated blobs are uncorrelated).

(ii) $C_\ell$ falls off at least as fast as $1/\ell^4$ for very large $\ell$, corresponding to scales much smaller than typical blob sizes (this follows from the simple assumption that the brightness is a *continuous* function of position).

(iii) If $\ell^2 C_\ell$ thus decreases both as $\ell$ gets small and as $\ell$ gets large, it must peak at some scale, a scale which we refer to as the coherence scale.

The behavior of the contaminant power spectrum for very small $\ell$ (whether there is indeed a coherence scale, *etc*), is of course quite a subtle one, as the presence of the Galactic plane means that the answer will be strongly dependent on which patches of sky are masked out during the analysis. These issues, as well as the effect of non-Gaussianity and inhomogeneity, are discussed in detail in [19]. In Figure 7, we have simply assumed that all three components have a coherence scale of about 10°, corresponding to $\ell \approx 10$, and used power spectra of the simple form $C_\ell \propto (5 + l)^{-3}$.

### 2.5.2. Frequency dependence

The frequency dependence of the three components has been extensively discussed in the literature (see *e.g.* [69] and references therein). For synchrotron radiation and free-free emission, Figure 7 uses simple power laws $B(\nu) \propto \nu^{-\beta}$. For synchrotron emission, $\beta \approx 0.75$ below 10 GHz [70], steepening to $\beta \sim 1$ above 10 GHz [71], so we simply assume $\beta = 1$ here. For free-free emission, we make the standard assumption $\beta = 0.1$. For dust, we assume a spectrum of the standard form

$$C_\ell \propto \frac{\nu^{3+\beta}}{e^{h\nu/kT} - 1}. \tag{7}$$

Although an emissivity index $\beta = 2$ is found to be a good fit in the Galactic plane [72], we use instead the more conservative parameters $T = 20.7$K, $\beta = 1.36$, which are found to better describe the data at high Galactic latitudes [69], since it is of of course the cleanest regions of the sky that are the most relevant ones for measurement of CMB fluctuations.

## 2.6. Local sources

In addition to the above-mentioned sources, all CMB experiments must of course tackle problems related to sidelobes from the sun, the moon, Earth and other planets, as well as electronic receiver noise. Ground- and balloon-based experiments will inevitable receive some atmospheric emission. On top of this, there are of course large numbers of potential sources of systematic errors. We will not comment further on most of these issues here, as they are often instrument-specific and must be modeled on a case-by-case basis.

The only local fluctuation source we will discuss here is receiver noise. Although we usually think of pixel noise as a problem of a different nature than the other contaminants, it has been shown [73,19] that it can in fact be described by an angular power spectrum $C_\ell(\nu)$ and so be treated on an equal footing. One may ask what the point is of doing this, since the statistical impact of the noise is straightforward to calculate anyway. The answer is that it provides better physical intuition. Real world brightness data is of course discretely sampled as "pixels" rather than smooth functions known at every point $\hat{\mathbf{r}}$, but as long as the sampling is sufficiently dense (the typical pixel separation being a few times smaller than the beamwidth), this discreteness is merely a rather irrelevant technical detail. It enters when we do the analysis in practice, but our results are virtually the same as if we had continuous sampling. In the common approximation that the beam profile is a Gaussian, one finds that the noise power spectrum is [73,19]

$$C_\ell = \frac{4\pi\sigma^2}{N} e^{\theta_b^2 \ell(\ell+1)}, \tag{8}$$

where $N$ is the number of pixels in the sky map, $\sigma$ is the r.m.s. pixel noise,

$$\theta_b \equiv \text{FWHM}/\sqrt{8\ln 2} \approx 0.425\,\text{FWHM}, \tag{9}$$

and FWHM is the full-width-half-max beam with. The so computed noise power for the proposed COBRAS/SAMBA satellite [74,19] is plotted in Figure 7, and should be fairly representative for next generation of spaceborne CMB missions.

## 3. The origin of primary anisotropies

In this section, we discuss the generation of primary CMB anisotropies in somewhat greater detail, with the emphasis on how the power spectrum depends on various cosmological parameters.

Traditionally (*e.g.* [75–78]), angular power spectra such as that in Figure 1 were computed with a "black box" approach: a gargantuan integration of the Boltzmann equation (see [79] and references in [9]), including all relevant physical processes, would be carried out numerically, and after an over-night run on a workstation, a power spectrum would result. Recently, great progress has been made [5,6,9,7,8,10] in providing an intuitive understanding of the inner workings of this black box, and it has been shown that power spectra accurate to within a few percent can be computed by much simpler means [8], by making certain physically motivated approximations. There are two distinct phases during which the calculations simplify considerably:

(i) Before recombination: acoustic oscillations

   As long as the hydrogen is mostly ionized, the mean free path of the CMB photons is so short that for all practical purposes, we can treat the photon-baryon plasma as a single fluid, which at each point has a well-defined velocity and density. This plasma will undergo acoustic oscillations until it recombines, and it has been shown [8] that these oscillations can be accurately treated with the WKB approximation.

(ii) After recombination: curvature and projection effects

   After recombination, things are complicated by the fact that the photons can stream freely and no longer have a well-defined velocity at each point. Rather, the evolution of their distribution in a 6-dimensional phase space must be computed. Fortunately, the fact that they are collisionless makes this quite simple, since we know that they will merely follow a null geodesic though spacetime until they reach us. In the end, we want the power spectrum as a function of multipole $\ell$ rather than wave number $\mathbf{k}$, and if space is flat, $C_\ell$ is given by an integral over $P(k)$ that has a simple geometrical interpretation. If space is positively or negatively curved ($\Omega + \Lambda \neq 1$), the geodesics will in addition converge or diverge so that the power spectrum in Figure 1 effectively shifts sideways, towards larger or smaller angular spaces, respectively.

The remainder of this section is largely based on the Ph.D. Thesis of Wayne Hu [9], to which the reader interested in more details is referred (it is available via anonymous ftp from *pac2.berkeley.edu*). Most of these details can also be found in [7,8].

## 3.1. Before recombination: acoustic oscillations

We will first turn to an old familiar example, that of classical perturbation theory in a single self-gravitating fluid. We will then present approximate results for the real-world case, where four components rather than one are present. After illustrating this with a simple toy model, we then discuss how the various real-world effects alter this simple solution and produce power spectra such as that in Figure 1.

### 3.1.1. The classical Jeans analysis

The classical equations governing a gas of density $\rho$, velocity $\mathbf{v}$ and pressure $p$ and the corresponding gravitational potential $\phi$ are the continuity equation, the Euler equation of motion and the Poisson equation of classical gravity, respectively:

$$\begin{cases} \dot{\rho} + \nabla \cdot (\rho \mathbf{v}) = 0 \\ \dot{\mathbf{v}} + (\mathbf{v} \cdot \nabla)\mathbf{v} = -\nabla(\phi + \frac{p}{\rho}) \\ \nabla^2 \phi = 4\pi G \rho \end{cases} \quad (10)$$

A simple solution to these equations is

$$\begin{cases} \rho_0(t, \mathbf{r}) = \frac{\rho_0}{a^3}, \\ \mathbf{v}_0(t, \mathbf{r}) = \frac{\dot{a}}{a}\mathbf{r}, \\ \phi_0(t, \mathbf{r}) = \frac{2\pi G \rho_0}{3}r^2, \end{cases}$$

where $a$ is a function of time that satisfies the Friedmann equation, and since the density is independent of $\mathbf{r}$, this classical solution clearly corresponds to the unperturbed FRW solution in General Relativity. From here on, let us write these fields as functions of the *comoving* position $\mathbf{x}$ rather than the physical position $\mathbf{r} = a(t)\mathbf{x}$. Let us now expand the fields as

$$\begin{cases} \rho(t, \mathbf{x}) = \rho_0(t)[1 + \delta(t, \mathbf{x})], \\ \mathbf{v}(t, \mathbf{x}) = \mathbf{v}_0(t, \mathbf{x}) + \mathbf{v}_1(t, \mathbf{x}), \\ \phi(t, \mathbf{x}) = \phi_0(\mathbf{x}) + \phi_1(t, \mathbf{x}), \end{cases}$$

and assume that $|\delta|$, $|\mathbf{v}_1|$, $\phi_1 \ll 1$. Substituting this into equation (10), dropping all non-linear terms, Fourier transforming everything with respect to $\mathbf{x}$, and doing some algebra, one obtains the second order ordinary differential equation

$$\ddot{\widehat{\delta}}(\mathbf{k}) + 2\frac{\dot{a}}{a}\dot{\widehat{\delta}}(\mathbf{k}) + \left(\frac{v_s^2|\mathbf{k}|^2}{a^2} - 4\pi G \rho_0\right)\widehat{\delta}(\mathbf{k}) = 0, \quad (11)$$

where hats denote Fourier transforms and $v_s$ is the sound speed. This well-known result illustrates two general features:

(i)  If $k < k_c \equiv (4\pi G\rho_0)^{1/2} a/v_s$, there will be a solution where the fluctuations grow without limit, whereas if $k > k_c$, the solutions will correspond to that of a damped harmonic oscillator.

(ii) The damping rate is given by $\dot{a}/a$, which we identify as the Hubble constant $H$.

*3.1.2. The relativistic multifluid case*

In the real world case, we need to consider relativistic effects, and four components instead of one: CDM, baryons, photons and neutrinos. Collisionless neutrinos of course cannot be modeled with an equation such as equation (11), since they do not have a unique velocity at each point. We will ignore them in our simplified discussion here and refer to [9,80,83] for details, since once the universe has become matter-dominated at $z \approx 25000\Omega_0 h^2$, massless or very light neutrinos will not contribute much to the total density. As mentioned above, before recombination, the baryons and the photons are so tightly coupled that we can treat them as one single fluid, thus leaving us with only two components: CDM and the photon-baryon plasma.

In a typical CDM scenario, the baryons constitute only a small fraction $\Omega_b/\Omega$ of the density of nonrelativistic matter, so after matter domination, equation (11) will accurately describe the CDM component if we merely set the pressure to zero ($v_s = 0$). After solving this equation, we can then take the resulting gravitational potential as given and obtain a self-contained equation for the evolution of the photon-baryon plasma. This is done in detail in [9] (§5.2), and the result is

$$\ddot{\hat{\Theta}} + \left(\frac{\dot{R}}{1+R}\right) H \dot{\hat{\Theta}} + (v_s k)^2 \hat{\Theta} = F, \qquad (12)$$

where dots denote derivatives with respect to the conformal time $\eta \equiv \int (1+z) dt$. The new variable $\Theta$ is simply defined to be a third of the density fluctuation, $\Theta = \delta/3$. This is more convenient, as this is the actual temperature fluctuation appearing in equation (1)†. The sound speed is given by

$$v_s = \frac{c}{\sqrt{3}} \frac{1}{\sqrt{1+R}}, \qquad (14)$$

---

† This is the origin of the factor 1/3: Since the photons are tightly coupled to the baryons, and the latter dominate the density, we have $n_\gamma \propto n_b \propto \rho_b \approx \rho$. Furthermore, blackbody radiation satisfies $n_\gamma \propto T^3$, so we have $T \propto \rho^{1/3}$. Hence the local temperature fluctuation is

$$\Theta \equiv \frac{\Delta T}{T} = \frac{1}{3}\frac{\Delta\rho}{\rho} = \frac{1}{3}\delta. \qquad (13)$$

where

$$R \equiv \frac{3\rho_b}{4\rho_\gamma} \approx \left(\frac{450}{1+z}\right)\left(\frac{h^2\Omega_b}{0.015}\right) \quad (15)$$

is, apart from a constant factor, just the baryon-to-photon density ratio. Note that as $z \to \infty$, $R \to 0$, so that the sound speed approaches the speed of light over $\sqrt{3}$.

Comparing equation (12) with equation (11), we notice that the self-gravity of the photon-baryon plasma (the $4\pi G$-term) has been neglected, with the assumption that $\Omega_b \ll \Omega$. Instead, the effects of gravity are all incorporated in the source term $F$ in equation (12). It describes the gravitational impact of the cold dark matter component, and in a full relativistic treatment [9] it is given by

$$F = \ddot{\widehat{\phi}}_c + \left(\frac{\dot{R}}{1+R}\right)H\dot{\widehat{\phi}}_c - \frac{k^3}{3}\widehat{\phi}. \quad (16)$$

Again, the $\phi$ to be used here is to a first approximation the peculiar gravitational potential produced by the CDM alone. We see that a second difference between equation (11) and equation (12) is the damping term, which now vanishes if the sound speed (and hence $R$) is time-independent. The quantity $\phi_c$ is the perturbation to the spatial curvature. This general-relativistic quantity has no Newtonian analog. To a first approximation, $\phi_c \approx -\phi$.

### 3.1.3. A toy model

To get a qualitative feeling for the solutions of equation (12), let us first solve a toy model based on the following crude approximations:

(i) The peculiar gravitational potential is time-independent, i.e., $\dot{\phi} = 0$.

(ii) The sound speed $v_s$ (and hence $R$) is independent of time.

The first approximation is valid in the matter-dominated ($z \ll 25000h^2$) linear regime, but clearly breaks down at very high redshifts. The second approximation is good as long as $R \ll 1$, i.e., long before recombination, when $v_s \approx c/\sqrt{3}$. With these approximations, equation (12) reduces to

$$\ddot{\widehat{\Theta}} + (v_s k)^2 \widehat{\Theta} = -\frac{k^2}{3}\widehat{\phi}. \quad (17)$$

We recognize this as the equation of motion for a simple harmonic oscillator with a time-independent driving force, and the solution is readily found to be

$$\widehat{\Theta}(\eta) = \left[\widehat{\Theta}(0) + (1+R)\widehat{\phi}\right]\cos(v_s k\eta) + \frac{1}{v_s k}\dot{\widehat{\Theta}}(0)\sin(v_s k\eta) - (1+R)\widehat{\phi}. \quad (18)$$

With the appropriate adiabatic initial conditions [9]

$$\begin{cases} \dot{\widehat{\Theta}}(0) = 0, \\ \widehat{\Theta}(0) = -\frac{2}{3}\widehat{\phi}, \end{cases} \quad (19)$$

the solution thus reduces to

$$\widehat{\Theta}(\eta) = \frac{\widehat{\phi}}{3}\cos(v_s k\eta) - \widehat{\phi} \quad (20)$$

very early on, when $R \approx 0$. Adiabatic initial conditions thus drive a cosine oscillation. In contrast, isocurvature initial conditions have $\widehat{\Theta}(0) = 0$ and drive a pure sine oscillation — we will not discuss the isocurvature case here, but refer to [9] for details.

### 3.1.4. *The random element*

On average, the random fields $\phi$, **v** and $\delta$ are of course all zero. Rather, the relevant quantities for our purposes are the variances, the power. Thus the contribution from the first and third terms in equation (1) is

$$V\left[\widehat{\phi} + \widehat{\Theta}\right] \equiv \left\langle \left|\widehat{\phi} + \widehat{\Theta}\right|^2 \right\rangle = \frac{1}{9}V[\widehat{\phi}]\cos^2(v_s k\eta). \quad (21)$$

As mentioned above, the fields in equation (1) were to be computed on the last-scattering surface, which means that we are only interested in the solutions at the fixed time $\eta = \eta_r$, the conformal time corresponding to recombination. Thus although we originally thought of equation (21) as an oscillating function of $\eta$, with $k$ a mere constant, we now think of it as function of $k$ instead, with $\eta$ held fixed. This is why the power spectrum in Figure 1 exhibits oscillatory behavior with respect to $\ell$.

As is discussed further on, although $\ell \asymp k$ in a certain sense, the correspondence is only approximate, which leads to an additional smearing of the oscillations.

### 3.1.5. *The Doppler peaks vanish...*

The "Doppler peaks" we derived above exhibit a qualitatively correct behavior when compared with Figure 1. If $\widehat{\phi}$ itself has a scale-invariant ("flat") spectrum, the solution of equation (21) will have a flat part for scales much smaller than the recombination horizon size (about 1°, corresponding to $\eta = \eta_r$), followed by a series of peaks. However, these peaks in fact have no Doppler in them, since we left out the Doppler term $\widehat{\mathbf{r}} \cdot \mathbf{v}$ from equation (1). Let us now remedy this. For the adiabatic case, the peculiar velocity is given by [9]

$$\widehat{\mathbf{v}}(\mathbf{k}) = -3\frac{\widehat{\mathbf{k}}}{k}\dot{\widehat{\Theta}} = (1+3R)v_s\widehat{\phi}\,\widehat{\mathbf{k}}\,\sin(v_s\eta k). \quad (22)$$

(The first equal sign follows from the continuity equation and irrotational flow.) Since the statistical properties of the random field are rotationally invariant, the mean squared radial velocity is simply $\langle |\hat{v}_r|^2 \rangle = \langle |\mathbf{v}|^2 \rangle/3$, so for $R = 0$, equation (22) gives

$$\langle |\hat{v}_r|^2 \rangle = \frac{1}{9} V[\hat{\phi}] \sin^2(v_s k \eta). \tag{23}$$

If we approximate the combined contribution from all three terms in equation (1) by adding equation (21) and equation (23) incoherently, we obtain

$$V[\Delta] = \frac{1}{9} V[\hat{\phi}] \left[ \sin^2(v_s \eta k) + \cos^2(v_s \eta k) \right] = \frac{1}{9} V[\hat{\phi}]. \tag{24}$$

In other words, when we included the Doppler contribution to the Doppler peaks, they vanished! It would thus appear as though we had made negative progress in this section.

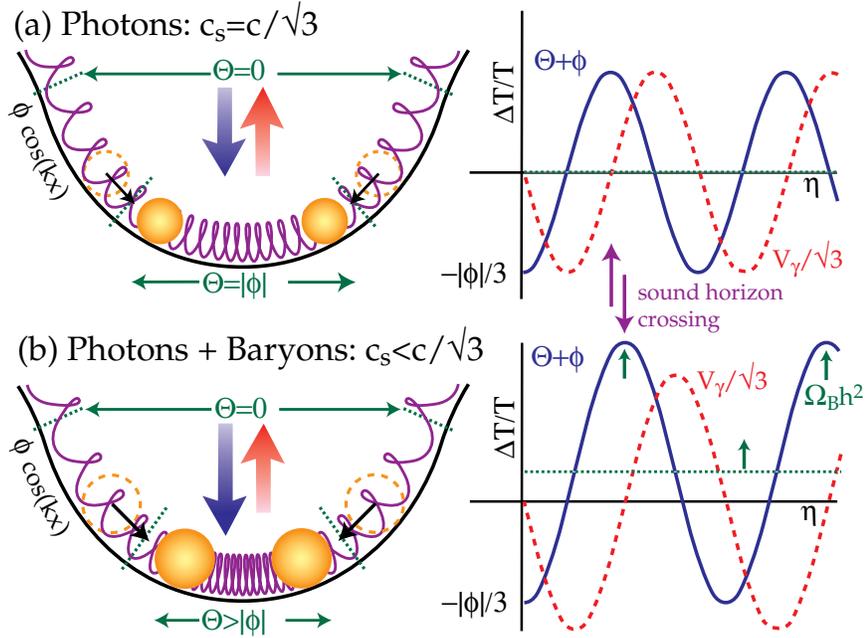

**Figure 8.** The origin of "Doppler peaks": why density and velocity oscillations do not cancel. Reprinted from [9].

### 3.1.6. ...and return

However, this cancellation occurs merely if the sound speed $v_S$ is exactly $c/\sqrt{3}$, i.e., if $R = 0$. This is illustrated in Figure 8, reprinted from [9]. In the right half of this figure, the solid curves correspond to the time-evolution of the combined density and gravity term, $\hat{\Theta} + \phi$, as given by equation (18). The dashed curve shows the time-evolution of the Doppler term, $v_r$, as given by equation (22). In the upper half, the sound speed is $v_s = c/\sqrt{3}$, so the solutions reduce to a cosine and a sine of the same amplitude, which give a mere constant when added in quadrature. When the sound speed is lower (which happens in reality, since $R$ grows with time according to equation (15)), two changes occur, both shown in the lower half, and both immediate consequences of equations (18) and (22):

(i) The amplitude of the dashed curve grows, but the amplitude of the solid curve grows much faster.

(ii) The solid curve gets shifted upwards, acquiring a non-zero offset.

In other words, the when the two curves are added in quadrature, the result will no longer be constant, but will exhibit bumps — the Doppler peaks have returned. The left-hand side of Figure 8 is an attempt to provide an intuitive understanding of what is happening in real space, in a potential well, for a single Fourier mode. The harmonic oscillator analogy is illustrated by two balls (whose distance from the top of the well represents $\widehat{\Theta}$, the overdensity in the photon-baryon fluid, and whose velocity represents the fluid velocity) connected by a spring (representing the pressure attempting to resist compression). Since we are assuming that $\phi$ is time-independent, the potential well itself will not change its shape over time, but both the density and the velocity of the matter in the well will oscillate. The adiabatic initial conditions (19) mean that initially, there is a slight overdensity inside the well ($\widehat{\Theta} = -(2/3)\widehat{\phi}$), with no fluid velocity at all ($\mathbf{v} \propto \dot{\widehat{\Theta}} = 0$). This corresponds to the balls being at rest at the dashed circles. Gradually, more matter starts falling into the well, reaching a maximum velocity when $\Theta = -\phi$ (the central value around which $\Theta$ is oscillating), until at a maximum density $\Theta = -(4/3)\phi$, the pressure manages to halt the compression and cause matter to begin streaming out of the potential well again. This maximum corresponds to the first "Doppler peak", which is often referred to as a compression peak. Note that this is a slightly confusing terminology, however, since it makes no sense if one forgets to specify where the compression occurs. This peak corresponds to maximum compression in the potential wells and maximum rarefaction at the potential peaks. At this point, $\widehat{\Theta} + \widehat{\phi}$ = $\widehat{\phi}/3$ in the upper plot. With the same terminology, the initial conditions ($t = \eta = 0$) corresponded to a rarefaction peak (maximum rarefaction in the wells, compression at the peaks), giving the Sachs-Wolfe contribution $\widehat{\Theta} + \widehat{\phi} = -\widehat{\phi}/3$.

The lower half of Figure 8 shows the real-world situation, where $v_s < c/\sqrt{3}$. Since $R > 0$, photons no longer dominate the fluid dynamics completely, so in the battle between pressure and gravity, the latter gains some territory compared with the $v_s = c/\sqrt{3}$ case. Compression will thus proceed further until pressure can reverse it (thus (i), the larger oscillation amplitude). Since the initial conditions are the same, not just the turning point, but also the zero-point of the oscillations (which is of course half-way down from the initial value to the turning-point), will shift. The result is (ii), *i.e.*, that the zero-point of oscillations in $\widehat{\Theta}$ exceeds $\phi$ in magnitude, so that there will on average be a blueshift from the potential well.

Effects (i) and (ii) are quite strong even for small changes in the sound speed. Plotting $[(\widehat{\Theta} + \widehat{\phi})^2 + v_r^2]$ using equations (18) and (22), lowering $v_s$ by merely 8% is sufficient to reproduce some features of Figure 1: the power spectrum: the result is flat for small $k$ and then rises to a Doppler peak about four times the height of the flat region. However, to make the toy model realistic, we must incorporate the fact that $R$ varies with time, as described below.

We conclude this section by noting that the locations of the peaks correspond to the maxima and minima of $\widehat{\phi} + \widehat{\Theta}$, the joint contributions of the gravity and density terms in equation (1), and *not* to those of the Doppler term $v_r$, whose contribution is zero at the peaks. In other words, the term "Doppler peaks" is strictly speaking a misnomer. However, since this term is so firmly entrenched in the CMB vocabulary, we will bow to convention and use it nonetheless.

### 3.1.7. *The WKB approximation*

Although the above toy model illustrated the mechanism by which the Doppler peaks are generated, it is not accurate enough to be useful in practice. One of the recent advances in this field was the realization [7,8] that if the WKB approximation is applied to equation (12), extremely accurate solutions can be obtained. The WKB approximation, often used in for instance quantum mechanics, reduces the solution of a second order linear ordinary differential equation to merely doing integrals, and works when the solution oscillates on a time scale much shorter than that on which the coefficients change. $R$ changes so slowly that this criterion is generally satisfied in equation (12), and the approximate solution is found to be [9]

$$[1 + R(\eta)]^{1/4} \widehat{\Theta}(\eta) = \widehat{\Theta}(0) \cos[kr_s(\eta)] + \frac{\sqrt{3}}{k} \left[ \dot{\widehat{\Theta}}(0) + \frac{1}{4}\dot{R}(0)\widehat{\Theta}(0) \right] \sin[kr_s(\eta)]$$
$$+ \frac{\sqrt{3}}{k} \int_0^\eta [1 + R(\eta')]^{3/4} \sin[kr_s(\eta) - kr_s(\eta')] F(\eta') d\eta', \quad (25)$$

where the *sound horizon* is defined as

$$r_s(\eta) = \int_0^\eta v_s(\eta') d\eta' = \frac{2}{3} \frac{1}{k_{eq}} \sqrt{\frac{6}{R_{eq}}} \ln\left[ \frac{\sqrt{1+R} + \sqrt{R+R_{eq}}}{1 + \sqrt{R_{eq}}} \right], \quad (26)$$

where $R_{eq}$ is $R$ at the time of matter-radiation equality, given by $z_{eq} \approx 24000\Omega_0 h^2$, and $k_{eq}$ is the wavenumber that enters the horizon at that time, given by $k_{eq} \approx (14\text{Mpc})^{-1}\Omega_0 h^2$. This approximation turns out to be so accurate that the weakest link becomes making an accurate model for the driving potential $F$.

## 3.2. During recombination: damping and diffusion

The above treatment was valid before recombination, when we could model the photon-baryon plasma as a single fluid. The fall-off of the power spectrum to the right in Figure 1 is due to the fact that recombination is not instantaneous, and caused by the following two effects:

(i) While the ionization fraction is so low that the photon mean free path is greater that the wavelength of a perturbation mode, but not yet so low that the photons have decoupled altogether, this mode will get suppressed. This is usually referred to as *photon diffusion* or *Silk damping* [84], and suppresses fluctuations in both photons and baryons.

(ii) When we measure the temperature in a given direction in the sky, we are averaging photons that last scattered near the front and near the back of the last scattering surface. This projection effect washes out fluctuations on scales smaller than the thickness of the last scattering surface.

## 3.3. After recombination: curvature and projection effects

So far, we have been discussing the evolution of 3D Fourier modes (plane waves). What we observe is of course a 2D temperature distribution on the celestial sphere. The result of converting the 3D results to a 2D power spectrum (see Appendix A for details about spherical harmonics) is that a spherical harmonic $a_{\ell m}$ is a weighted average of the 3D fluctuations on many different scales. Although for a given multipole $\ell$, the weights (sometimes referred to as the *window function*) tend to peak at a characteristic wavenumber $k \propto \ell$, it has a non-negligible width. The effect of this is essentially that the angular power spectrum will be a *smeared out* version of the 3D power spectrum, with all sharp features slightly softened. In addition, the proportionality constant between $k$ and $\ell$ depends on the curvature of the space, increasing with $\Omega_0$. In other words, if we decrease $\Omega_0$, the angular power spectrum of Figure 1 will basically shift to the right, since the same physical scale on the last scattering surface will subtend a smaller angle in the sky. Although these physical scales of course also depend slightly on $\Omega_0$, and not all in the same way, the overall scaling is roughly that given by equation (3), *i.e.*, $\ell \propto \Omega_0^{-1/2}$.

## 3.4. Parameter dependence

Let us conclude our discussion of Doppler peaks by summarizing what we care the most about: how the details of the power spectrum depend on various cosmological parameters.

$\Omega_0$ As mentioned above, lowering $\Omega_0$ primarily shifts the power spectrum to the right due to the curvature effect. In addition, the lowest multipoles get boosted by the late ISW effect.

$\Omega_b$ As we saw in section 3.1.5, the power spectrum had no peaks if $R = 0$. Since $R \propto h^2 \Omega_b$, increasing the baryon fraction will thus make the peaks higher. In addition, as was illustrated in Figure 8, this boosts the compression peaks (number 1, 3, *etc.*) much more than the rarefaction peaks (number 2, 4, *etc.*). This effect is illustrated in Figure 9.

$h$ The dependence on the Hubble constant when $h^2\Omega_b$ is held fixed at the nucleosynthesis value $h^2\Omega_b = 0.015$ is shown in Figure 10. Potential decay due to radiation pressure inside the horizon during radiation domination boosts the peaks via the $F$-term, so lowering $h$ will increase this boost by delaying matter-radiation equality. By changing the expansion rate, this also shifts the peaks somewhat.

$\Lambda$ Increasing the cosmological constant while keeping space flat ($\Omega_0 + \Lambda = 1$) will among other things boosts the lowest multipoles via the late ISW effect.

$n$ Increasing $n$, the tilt of the primordial power spectrum of $\phi$, will of course increase the slope of the angular power spectrum $C_\ell$ as well, raising the right side relative to the left side.

$\tau$ Early reionization will, as discussed in section 2.2.3 suppress the power at $\ell \gg 10$.

Some of these effects are illustrated in Figure 2. Here an arrow next to a quantity indicates the direction in which the curve with the same line type as the arrow will shift if the parameter is increased.

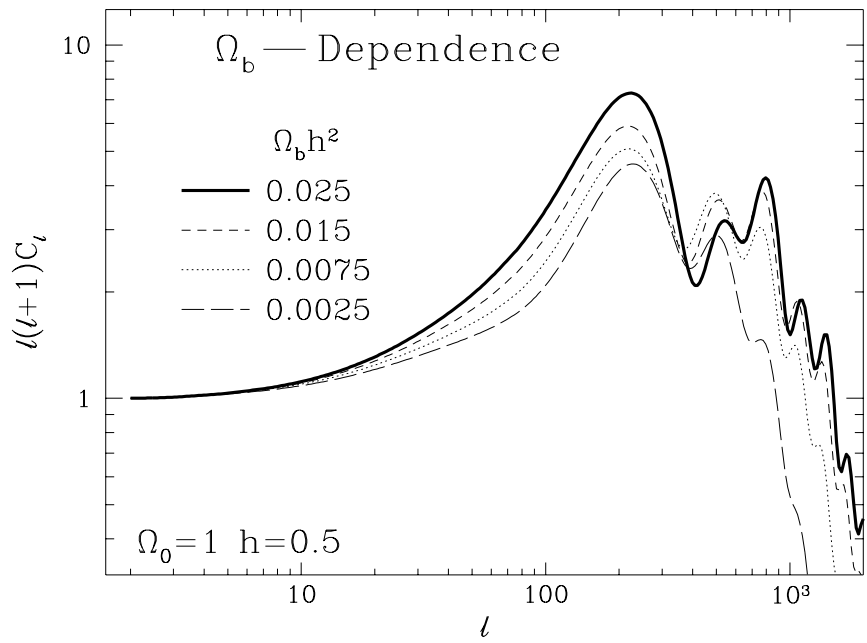

**Figure 9.** The effect of changing the baryon fraction $\Omega_b$. Reprinted from [9].

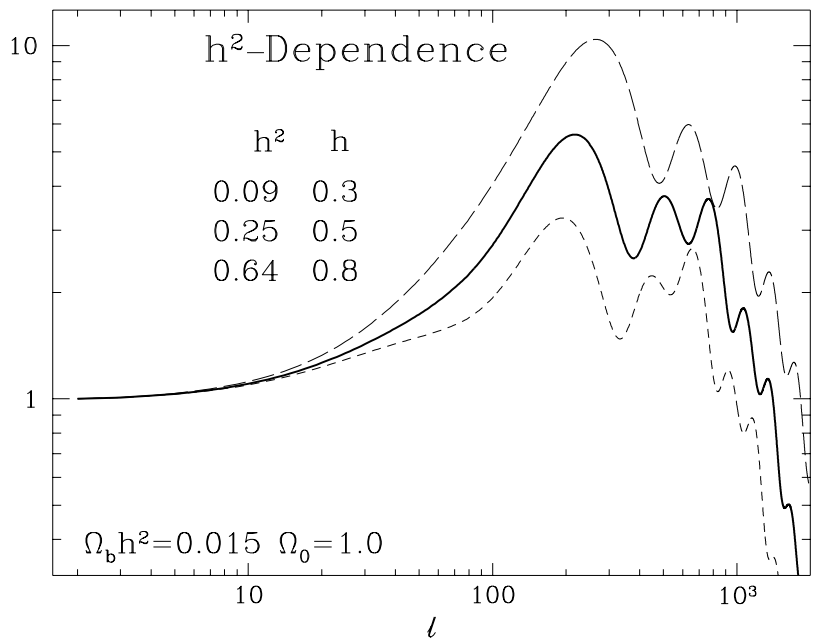

**Figure 10.** The effect of changing the Hubble constant. From top to bottom, the curves correspond to $h = 0.3, 0.5$ and $0.8$, respectively. Reprinted from [9].

# 4. How accurately can the CMB pin down cosmological parameters?

Above we have discussed the various fluctuation sources listed in Table 1. It was seen that power spectra such as that in Figure 1 depend on virtually all key cosmological parameters, which means that accurate measurement of the power spectrum should enable us to make estimates of all these parameters. In this section, we discuss the usefulness of the CMB for parameter estimation, and describe how to assess the accuracy with which the different parameters can be measured. For additional discussion of these issues, see [85,73,86].

## 4.1. The parameter estimation problem

Let us define $\widehat{a}_{\ell m}$ as the the observed multipole coefficients (including both the noise contribution and cosmic variance). This is a priori a vector of random variables with zero mean, $\langle a_{\ell m} \rangle = 0$. Let us for the moment ignore the complication of the galactic plane, and assume that we have complete sky coverage. The covariance matrix will then be diagonal, and, using equation (8), is given by

$$\langle \widehat{a}_{\ell m}^* \widehat{a}_{\ell m} \rangle = \delta_{\ell\ell'} \delta_{mm'} C_\ell^{\text{tot}}, \tag{27}$$

where

$$C_\ell^{\text{tot}} \equiv C_\ell + \frac{4\pi\sigma^2}{N} e^{\theta_b^2 \ell(\ell+1)}. \tag{28}$$

Here $C_\ell$ denotes the true CMB power spectrum, which depends on some set of cosmological parameters that we will denote $\boldsymbol{\Theta} = (\theta_1, \theta_2, ..., \theta_n)$. We might for instance consider an $n$-parameter model where $n = 9$ and the parameters are

$$\boldsymbol{\Theta} = (Q, \Omega_0, \Omega_b h^2, h, n, \Lambda, T/S, \tau); \tag{29}$$

the quadrupole normalization, the density parameter, the baryon density, the Hubble parameter, the spectral index, the cosmological constant, the ratio of tensor to scalar modes at $\ell = 2$ and the optical depth due to reionization, respectively.

Suppose that we have measured $\widehat{a}_{\ell m}$ for all $\ell$ up to $\ell_{\max}$. The standard way to estimate these $n$ parameters simultaneously from the observed multipole coefficients $\widehat{a}_{\ell m}$ is to compute the maximum likelihood (ML) estimate, the value of the vector $\boldsymbol{\Theta}$ that maximizes the likelihood function $L(\widehat{a}_{\ell m}, \boldsymbol{\Theta})$. This is of course equivalent to minimizing the quantity $\mathcal{L} \equiv -2 \ln L$. $L$ is simply the probability distribution for the random vector $\widehat{a}_{\ell m}$ given the model parameters $\boldsymbol{\Theta}$. With the standard assumption that the random fields are Gaussian, $\widehat{a}_{\ell m}$ is just a vector of independent Gaussian random variables with variances given by equation (27), and dropping an irrelevant additive

constant, we obtain

$$\mathcal{L} = \sum_{\ell=2}^{\ell_{\max}} (2\ell + 1) \left[ \ln C_\ell + \frac{\widehat{C}_\ell}{C_\ell} \right] \tag{30}$$

where the observed power spectrum $\widehat{C}_\ell$ is defined as

$$\widehat{C}_\ell \equiv \frac{1}{2\ell + 1} \sum_{m=-\ell}^{\ell} |\widehat{a}_{\ell m}|^2. \tag{31}$$

### 4.2. The error bars

Since $\widehat{a}_{\ell m}$ is a random variable, so is the estimated parameter vector $\Theta$. A convenient measure of the resulting errors is thus the covariance matrix of $\Theta$, defined as

$$M \equiv \langle (\Theta - \Theta_0)(\Theta - \Theta_0) \rangle, \tag{32}$$

where $\Theta_0 \equiv \langle \Theta \rangle$. If the ML-estimates are unbiased, then $\Theta_0$ equals the true parameter values. If not, the bias is normally corrected for with Monte-Carlo simulations.

Let us Taylor expand $\mathcal{L}$ around the ML-estimate $\Theta$. By definition, all first derivatives $\partial \mathcal{L}/\partial \theta_i$ will vanish at the ML-point, so that the local behavior will be dominated by the quadratic terms. Since $L = \exp[-\mathcal{L}/2]$, we see that the likelihood function will be approximately Gaussian near the ML-point. If the error bars are quite small, $L$ usually drops sharply before third order terms have become important, so that this Gaussian is a good approximation to $L$ everywhere. The covariance matrix $M$ is then given simply by the second derivatives at the ML-point, as the inverse of the Hessian matrix: $M = I^{-1}$, where

$$I_{ij} \equiv \frac{\partial^2 \mathcal{L}}{\partial \theta_i \partial \theta_j}. \tag{33}$$

In statistics, $\langle I \rangle$ is known as the Fisher information matrix, and it can be proved that $\langle I \rangle^{-1}$ in a certain sense gives the smallest error bars that can possibly be extracted from the data, regardless of whether ML-estimation or some other parameter fitting method is used [87,88]. A straightforward calculation shows that at the ML-point,

$$I_{ij} \equiv \frac{\partial^2 \mathcal{L}}{\partial \theta_i \partial \theta_j} = \sum_{\ell=2}^{\ell_{\max}} (2\ell + 1) \left[ C_\ell + \frac{4\pi \sigma^2}{N} e^{\theta_b^2 \ell(\ell+1)} \right]^{-2} \left( \frac{\partial C_\ell}{\partial \theta_i} \right) \left( \frac{\partial C_\ell}{\partial \theta_j} \right). \tag{34}$$

This handy expression, also given by [86], tells us that the crucial functions which determine the attainable accuracy are the *derivatives* of the power spectrum with respect to the various parameters. Examples of such derivatives are shown in Figure 11, and the reader is encouraged to try to interpret the behavior or these curves in terms of our discussions of parameter dependence in the previous sections. For instance, $\partial C_\ell / \partial \tau$ is shaped as $-C_\ell$ for $\ell \gg 10$, since earlier reionization suppresses all these multipoles by the same factor $\varepsilon^{-2\tau}$. $\partial C_\ell / \partial Q$ of course has the same shape as the power spectrum itself.

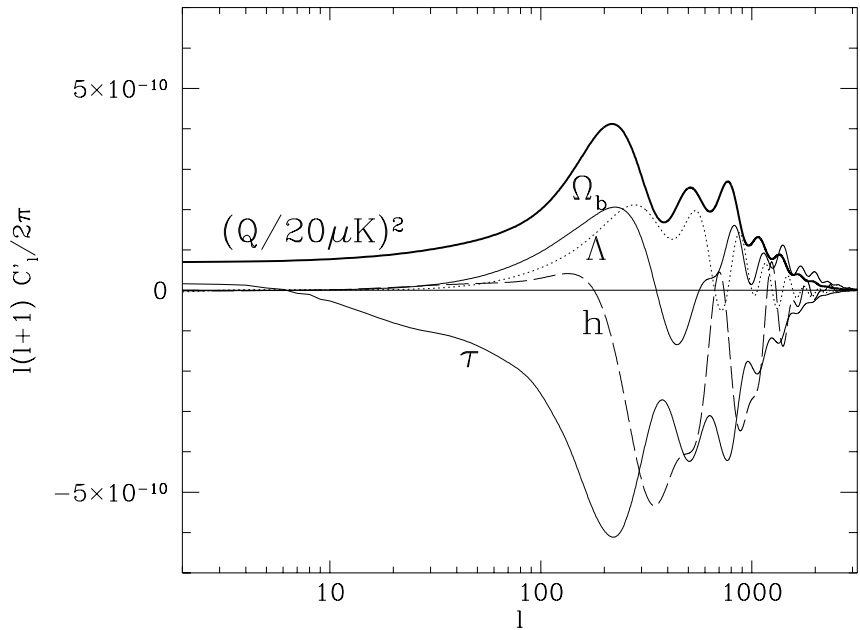

**Figure 11.** The derivatives of the CDM power spectrum with respect to various parameters.

### 4.3. A geometric interpretation

Equation (34) has a simple geometric interpretation. We can think of the $n$ functions $\partial C_\ell/\partial\theta_i$ as vectors in a Hilbert space of dimension $(\ell_{\max} - 1)$, and think of $I_{ij}$ as simply the dot product of the vectors $\partial C_\ell/\partial\theta_i$ and $\partial C_\ell/\partial\theta_j$. This dot product gives a rather small weight to low $\ell$-values, essentially because there are fewer $m$-modes there and correspondingly larger cosmic variance. In addition, the weights are seen to be exponentially suppressed for large $\ell$, where beam dilution causes the effect pixel noise to explode. If the parameter dependence of $C_\ell$ was such that all $n$ vectors $\partial C_\ell/\partial\theta_i$ were orthogonal under this dot product, then $I$ and $M$ would be diagonal, and the errors in the estimates of the different parameters would be uncorrelated. The more similarly shaped two curves in Figure 11 are, the harder it will be to break the degeneracy between the corresponding two parameters. In the extreme case where two curves have identical shape (are proportional to each other), the matrix $I$ becomes singular and the resulting error bars on the two parameters become infinite. It is therefore interesting to diagonalize the matrix $M$ (or equivalently, its inverse $I$). The eigenvectors will tell us which $n$ parameter combinations can be independently estimated from the CMB data, and the corresponding eigenvalues tell us what how accurately this can be done.

It has been pointed out [85] that there will be a considerable parameter degeneracy if data is only availably up to around the first Doppler peak. This is clearly illustrated in Figure 11: most of the curves lack strong features in this range, so some of them can be well approximated by linear combinations of the others. If a CMB experiment has high enough resolution to measure the power out to $l \sim 10^3$, however, this degeneracy is broken, and for the 9-parameter model we discussed, most parameters can be estimated to an accuracy of a few percent or better assuming typical noise levels. This agrees well with the results of [86], who compute the attainable error bars on $\Omega_0$, and [73] where the attainable error bars on $n$ and $T/S$ are estimated. Even the abundance of hot dark matter can be constrained in this fashion [83].

In should be stressed that the formalism above only tells us what the attainable accuracy is if the truth is somewhere near the point in parameter space at which we compute the power derivatives. Figure 11 corresponds to standard COBE-normalized CDM, *i.e.*, to $\boldsymbol{\Theta} = (20\mu K, 1, 0.015, 0.5, 1, 0, 0, 0)$. The worst scenario imaginable would probably be extremely early reionization, since $\tau \gg 1$ would eliminate almost all small-scale power and introduce a severe degeneracy by making all the power derivatives almost indistinguishable for $\ell \gg 10$, the region where they differ the most in Figure 11.

Needless to say, accurate parameter estimation also requires that we can compute the theoretical power spectrum accurately. It has been argued [80] that this can be done to better than 1%, by accurately modeling various weak physical effects such as Helium recombination and polarization.

### 4.4. Real-world complications: foregrounds and incomplete sky coverage

Above we ignored two ubiquitous real-world problems: foregrounds will always be present at some level, and all-sky data will not be available (because of the hopeless foreground problems in the galactic plane).

#### 4.4.1. Foregrounds

From the structure of equation (34), the impact of residual foreground contamination (or the presence of any other fluctuation source of known amplitude) is clear: the foreground power $C_\ell$ simply gets added to the expression in square brackets. For instance, strong foreground fluctuations around $\ell = 300$ would reduce the weight given to those $\ell$-values in the dot product. Thus even if the various curves in Figure 11 look quite different there, these differences would carry little statistical weight, degeneracy problems would increase and error bars on the parameter estimates would grow.

Estimates of residual foreground contamination are shown in Figure 12. They are based on the foreground models of section 2.3 and the specifications of the proposed 9-channel COBRAS/SAMBA satellite, and are further described in [19]. The top curve (shaded) is a standard CDM power spectrum. From top to bottom, the remaining curves show the residuals obtained with four different subtraction schemes, of increasing levels of sophistication. The uppermost one corresponds to no subtraction whatsoever, and using observations in merely a single frequency channel, 143 GHz. The second one also involves no subtraction, but using the channel that has the smallest total foreground signal at each multipole. The optimal frequency as a function of $\ell$ is shown by the heavy solid line in Figure 7, and is seen to increase with frequency, reflecting the geography of the $\ell - \nu$ plane. The third curve (dashed) is based on the multi-frequency subtraction scheme of [19], applied on a pixel-by-pixel basis. It essentially involves fitting for the various foreground components using the data at all frequencies, but in addition takes noise levels into account to minimize the r.m.s. of the residuals. The fourth curve uses the same method [19], but applied in Fourier space instead of in real space, *i.e.*, subtracting mode by mode rather than pixel by pixel. By taking advantage of the $\ell$-dependence of the foregrounds, this is seen to reduce the residuals by another factor of 10. If our foreground models turn out to be roughly correct, the results are of course quite encouraging. Since the power is the square of the quantity on the vertical axis, even the extremely simple-minded "best channel" approach would give $C_\ell$-errors at the 1% level, and it seems quite plausible that subtraction methods will allow us to reduce foreground contamination to negligible levels all the way out past the first few Doppler peaks.

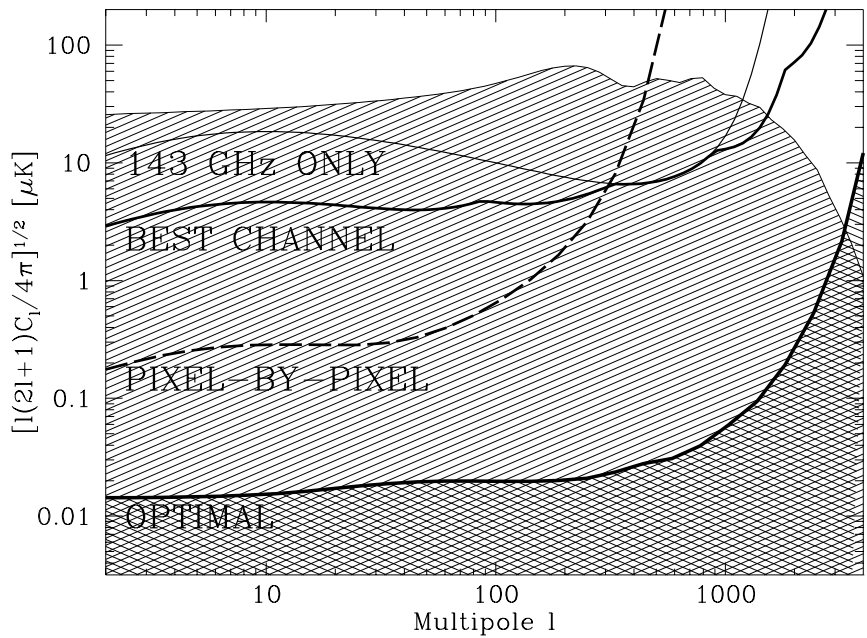

**Figure 12.** The residual foreground contamination with various subtraction schemes.

*4.4.2. Incomplete sky coverage*

The reason that equation (34) has such a simple form is that the covariance matrix in equation (27) is diagonal. This happens only if we have complete sky coverage. Even if a CMB experiment produces an all-sky CMB map, the presence of "dirty" regions such as the Galactic plane, the Large Magellanic Cloud, bright point sources, *etc.*, means that we may want to throw away some of the pixels, leaving us with a map with a topology reminiscent of a Swiss cheese. Incomplete sky coverage impedes parameter estimation in two different ways, related to sample variance and spectral smearing.

Firstly, the **sample variance increases**, since we are effectively probing fewer independent modes. The sample variance essentially grows as [60] $\sqrt{4\pi/\Omega}$, where $\Omega$ is the solid angle covered by the map, so this effect will basically increase the $C_\ell$-term in square brackets in equation (34) by this factor.

Secondly, it is well-known that it is impossible to compute the exact coefficients $a_{\ell m}$ using merely part of the sky. Instead of the spherical harmonics, we must expand our maps in some other set of basis functions, functions that vanish in all the "holes" in our maps. In contrast to the spherical harmonics, each of these functions will inevitably probe a range of $\ell$-values, rather than just a single multipole, specified by a *window function*. The result will be that the functions in Figure 11 will have their **features smeared out** on scales smaller than the window function width, thus reducing the differences between them and increasing the parameter degeneracy. Because of this, we clearly want these window functions to be as narrow as possible.

A prescription for how to calculate such basis functions, taking incomplete sky coverage, pixelization, and position-dependent noise into account, is given in [16], and it is found that given a patch of sky whose smallest angular dimension is $\Delta\theta$, each basis function will probe an $\ell$-band of width $\Delta\ell \approx 60°/\Delta\theta$. For instance, if we restrict our analysis to a 10° × 10° square, then $\Delta\ell \approx 6$. This is very good news. It means that the only performance degradation in our parameter estimation will stem from the fact that we are unable to take advantage of sharp features in the power spectra of width $\Delta\ell \approx 6$ or smaller. This is essentially no loss at all, since all mainstream models predict fairly smooth power spectra, without any sharp spikes or discontinuities.

# 5. Summary

We have attempted to summarize the many sources of CMB anisotropies, ranging from useful primary and secondary sources that probe cosmological parameters to "tertiary" foreground sources that are an impediment to parameter estimation. We have also described how to assess the accuracy with which these parameters can be measured. We summarize the current situation as follows.

(i) The CMB power spectrum depends sensitively on virtually all key cosmological parameters, which means that future CMB experiments have the potential to measure these parameters with hitherto unprecedented accuracy.

(ii) Galactic and extragalactic foregrounds tend to differ from the CMB not only in their frequency spectra but also in their power spectra, as summarized in Figure 7. We can take advantage of this to subtract them out more accurately.

(iii) By combining high resolution CMB maps from a wide range of frequencies, it appears feasible to reduce residual foreground contamination to negligible levels in a substantial fraction of the sky.

(iv) Incomplete sky coverage is not likely to substantially reduce our ability to constrain cosmological parameters. It merely increases the sample variance somewhat and eliminates our ability to take advantage of sharp features in the power spectrum — and none of the mainstream models predict such sharp features.

(v) Within a CDM framework, the only scenario that would ruin our ability to measure cosmological parameters is extremely early reionization. Fortunately, recent CMB data on degree scales appear to rule out this possibility [3].

In conclusion, it appears as though CMB workers have reason to view the future with mild optimism.

*Acknowledgements* I would like to thank Wayne Hu, Lloyd Knox, Douglas Scott and Uroš Seljak for helpful comments on the manuscript. Wayne and Uroš kindly provided figures, and Naoshi Sugiyama supplied power spectrum data. I would also like to thank the organizers of this school, Silvio Bonometto and Joel Primack, for an enjoyable and productive meeting.

## Appendix A.   Cosmic statistics — a primer

In this appendix, we review some of the mathematical underpinnings of CMB modeling — random fields, power spectra, spherical harmonics, *etc.*

### A.1. Random fields

The mathematical theory of random fields (sometimes known as three-dimensional stochastic processes) is a very useful tool when analyzing cosmological structure formation. A *random field* is simply an infinite-dimensional random variable, such that each realization of it is a field on some space. In cosmology applications, these fields tend to be $\delta$, $\mathbf{v}$ or $\phi$, and the space is physical space at some fixed time $t$ (or in the CMB case, the surface of the celestial sphere). As every quantum field theorist knows, it is a nightmare to try to define a nice measure on an infinite-dimensional space, so random fields are defined by specifying the joint probability distribution of their values at any $n$ points, $n = 1, 2, 3, ...$, thus circumventing the need to define a probability distribution on the space of all fields. Hence, to define a random field $\delta$, one must specify the 1-point distribution (the 1-dimensional probability distribution of $\delta(\mathbf{x}_1)$ for all $\mathbf{x}_1$), the 2-point-distribution (the 2-dimensional probability distribution of the vector $[\delta(\mathbf{x}_1), \delta(\mathbf{x}_2)]$ for all $\mathbf{x}_1$ and $\mathbf{x}_2$), *etc.* In cosmology, the random fields are always assumed to be translationally and rotationally invariant, *i.e.* homogeneous and isotropic. Hence the 1-point distribution is independent of $\mathbf{x}_1$, and the 2-point distribution will depend only on the scalar quantity $x \equiv |\mathbf{x}_2 - \mathbf{x}_1|$.

*A.1.1. Ergodicity*

A random field $\delta$ is said to be *ergodic* if one can use ensemble averaging and spatial averaging interchangeably. The ensemble average of a random field $\delta$ at a point, denoted $\langle \delta(\mathbf{x}) \rangle$, is simply the expectation value of the random variable $\delta(\mathbf{x})$. Thus for an ergodic field,

$$\langle \delta(\mathbf{x}_1) \rangle = \lim_{R \to \infty} \left(\frac{4}{3}\pi R^3\right)^{-1} \int_{|\mathbf{x}|<R} \delta_*(\mathbf{x}) d^3x$$

holds for all points $\mathbf{x}_1$ and for all realizations $\delta_*(\mathbf{x})$ (except for a set of probability measure zero). Ensemble averages are completely inaccessible to us, since we have only one universe to look at, namely the particular realization that we happen to live in. So as cosmologists, we are quite happy if we have ergodicity, since this means that we can measure these elusive ensemble averages by simply averaging over large volumes instead.

*A.1.2. Gaussianity*

A random field is said to be *Gaussian* if all the above-mentioned probability distributions are Gaussian. This is a very popular assumption in cosmology, partly because, as we will see, it greatly simplifies matters. A first nice feature of this assumption is that all homogeneous and isotropic Gaussian random fields are ergodic†. Taking the spatial average of the definition of $\delta$, for instance, ergodicity implies that

$$\langle \delta(\mathbf{x}) \rangle = 0.$$

Let us define the *correlation function* as

$$\xi(x) \equiv \langle \delta(\mathbf{x}_2)\delta(\mathbf{x}_1) \rangle.$$

(Note that because of the homogeneity and isotropy, the correlation function depends only on $x \equiv |\mathbf{x}_2 - \mathbf{x}_1|$.) Since the $n$-point distribution is Gaussian, it is defined by its mean vector $\langle \delta(\mathbf{x}_n) \rangle$ (which is identically zero) and its covariance matrix

$$C_{mn} \equiv \langle \delta(\mathbf{x}_m)\delta(\mathbf{x}_n) \rangle = \xi(|\mathbf{x}_m - \mathbf{x}_n|).$$

Thus the Gaussian random field $\delta$ has the extremely useful property that it is is entirely specified by its correlation function.

## A.2. The 3D power spectrum

If we Fourier expand $\delta$ as

$$\delta(\mathbf{x}) = \frac{1}{(2\pi)^3} \int e^{i\mathbf{k}\cdot\mathbf{x}} \widehat{\delta}(\mathbf{k}) d^3k,$$

---

† Note that this is true only for random fields that live on infinite spaces such as $\mathsf{R}^n$, and does not hold for fields on compact manifolds such as the sphere $S^2$. Thus the field of microwave background anisotropies (defined further on) is not ergodic, so that even if we could reduce our experimental errors to zero, we could still never measure any ensemble-averages with perfect accuracy. This phenomenon is known as "cosmic variance". It stems from the fact that in the spatial average above, one cannot average over an infinite volume (*i.e.* let $R \to \infty$), since the volume of the compact manifold (in this case the area of the sphere) is finite.

we see that if its Fourier transform

$$\widehat{\delta}(\mathbf{k}) = \int e^{-i\mathbf{k}\cdot\mathbf{x}}\delta(\mathbf{x})d^3x,$$

is a Gaussian random variable for any $k$, then $\delta$ will automatically be a Gaussian random field (since a sum of Gaussians is always Gaussian). Cosmologists like to postulate that the complex numbers $\widehat{\delta}(\mathbf{k})$ have *random phases*, which implies that they are all uncorrelated. One postulates that

$$\langle\widehat{\delta}(\mathbf{k})^*\widehat{\delta}(\mathbf{k}')\rangle = (2\pi)^3\delta(\mathbf{k}-\mathbf{k}')P(\mathbf{k}),$$

where the function $P(\mathbf{k})$ is called the *power spectrum* and $\delta$ is the Dirac delta function (which will hopefully not be confused with the random field $\delta$). This implies that even if $\widehat{\delta}(\mathbf{k})$ does not have a Gaussian distribution, the random field $\delta$, being an infinite sum of independent random variables, will still be Gaussian by the Central Limit Theorem for many well-behaved power spectra. It is straightforward to show that the power spectrum is simply the three-dimensional Fourier transform of the correlation function, *i.e.*

$$P(k) = 4\pi\int\left(\frac{\sin kr}{kr}\right)\xi(r)r^2 dr.$$

Note that $P$ depends on $\mathbf{k}$ only through its magnitude $k = |\mathbf{k}|$, because of the isotropy assumption.

### A.3. Working with spherical harmonics

For functions that live on a sphere, the analogue of a Fourier expansion is an expansion in *spherical harmonics*. The spherical harmonics are defined as

$$Y_{\ell m}(\theta,\varphi) = \sqrt{\frac{2\ell+1}{4\pi}\frac{(\ell-m)!}{(\ell+m)!}}P_\ell^m(\cos\theta)e^{im\varphi},$$

where $P_\ell^m$ are the associated Legendre functions, and $\ell$ and $m$ are integers such that $\ell \geq 0$ and $|m| \leq \ell$. They have the symmetry property that

$$Y_{\ell,-m} = (-1)^m Y_{\ell m}^*,$$

where $*$ denotes complex conjugation. These functions form a complete orthonormal set on the unit sphere. The orthonormality means that

$$\int Y_{\ell m}^*(\theta,\varphi)Y_{\ell' m'}^*(\theta,\varphi)d\Omega = \delta_{\ell\ell'}\delta_{mm'},$$

and the completeness means that we can expand any $L^2$ function $\Delta$ as

$$\Delta(\theta, \varphi) = \sum_{\ell=0}^{\infty} \sum_{m=-\ell}^{\ell} a_{\ell m} Y_{\ell m}(\theta, \varphi),$$

where

$$a_{\ell m} \equiv \int Y_{\ell m}^*(\theta, \varphi) \Delta(\theta, \varphi) d\Omega.$$

We use the differential solid angle notation

$$d\Omega = \sin\theta \, d\theta d\varphi.$$

Also, we will sometimes replace $\theta$ and $\varphi$ by the unit vector

$$\hat{\mathbf{n}} = (\sin\theta \cos\varphi, \sin\theta \sin\varphi, \cos\theta)$$

and write things like

$$Y_{\ell m}(\theta, \varphi) = Y_{\ell m}(\hat{\mathbf{n}}).$$

With this notation, the so called addition theorem for spherical harmonics states that

$$\sum_{m=-\ell}^{\ell} Y_{\ell m}^*(\hat{\mathbf{n}}) Y_{\ell m}(\hat{\mathbf{n}}') = \frac{2l+1}{4\pi} P_\ell(\hat{\mathbf{n}} \cdot \hat{\mathbf{n}}'),$$

where $P_\ell = P_\ell^0$ are the Legendre polynomials.

For the reader who wants more intuition about spherical harmonics, it is good to lump together all harmonics with the same $\ell$ value. For $\ell = 0$, 1, 2, 3 and 4, these sets are referred to as the monopole, the dipole, the quadrupole, the octupole and the hexadecapole, respectively. Geometrically, when one expands a function $f(\hat{\mathbf{n}})$ in spherical harmonics, $\ell = 0$ picks up the constant part, $\ell = 1$ picks up the remaining linear part, $\ell = 2$ picks up the remaining quadratic part, $\ell = 3$ picks up the remaining cubic part, *etc*. In group theory jargon, the spherical harmonics corresponding to different $\ell$-values are the irreducible representations of the group of rotations of the sphere. This means that if one expands a rotated version of the same function $f$, the new spherical harmonic coefficients $a'_{\ell m}$ will be some linear combination of the old ones $a_{\ell m}$,

$$a'_{\ell m'} = \sum_{m=-\ell}^{\ell} D_{\ell m' m} \, a_{\ell m},$$

such that different $l$-values live separate lives and never mix. For instance, however one chooses to rotate a linear function ($\ell = 1$), it will always remain linear and never say pick up quadratic terms.

## A.4. Random fields on the sphere

$T_\gamma(\hat{\mathbf{n}})$, the CBR temperature that we observe in the direction $\hat{\mathbf{n}}$ in the sky, is modeled as a random field. It is more convenient to work with the dimensionless version

$$\Delta(\hat{\mathbf{n}}) \equiv \frac{T(\hat{\mathbf{n}})}{\langle T(\hat{\mathbf{n}})\rangle} - 1,$$

which is often denoted $\Delta T/T$. All the formulas and definitions we gave for $\delta$ have spherical analogues for $\Delta$, as we will now see. The *angular correlation function* is defined as

$$c(\theta) = \langle \Delta(\hat{\mathbf{n}})\Delta(\hat{\mathbf{n}}')\rangle,$$

where $\hat{\mathbf{n}}\cdot\hat{\mathbf{n}}' = \cos\theta$, and the right hand side is independent of the actual directions by the isotropy assumption. It is easy to show that this implies *random phases*, that

$$\langle a^*_{\ell m} a_{\ell' m'}\rangle = \delta_{\ell\ell'}\delta_{mm'} C_\ell$$

for some coefficients $C_\ell$. These coefficients $C_\ell$ constitute the spherical version of the power spectrum $P(k)$, and are usually referred to as the *angular power spectrum*. Just as the spatial correlation function was the Fourier transform of the spatial power spectrum, the angular correlation function is what might be called a "Legendre transform" of the angular power spectrum $C_\ell$. Using the addition theorem, one readily obtains

$$c(\theta) = \frac{1}{4\pi}\sum_{\ell=0}^{\infty}(2l+1)C_\ell P_\ell(\cos\theta).$$